\documentclass[aps,pra,twocolumn,superscriptaddress]{revtex4}
\usepackage{mathtools}
\usepackage{graphicx}
\usepackage{epstopdf}
\usepackage{lmodern}
\usepackage{xfrac}
\usepackage{natbib}
\usepackage[breaklinks]{hyperref}
\usepackage{verbatim}

\usepackage{mathtools}
\DeclarePairedDelimiter\ceil{\lceil}{\rceil}

\newcommand{\ket}[1]{|{#1}\rangle}
\newcommand{\bra}[1]{\langle{#1}|}

\newcommand{\norm}[1]{\left\lVert #1 \right\rVert}

\newcommand{\diff}{\mathrm{d}}
\newcommand{\binwidth}{\tau}

\newcommand{\epsc}{\epsilon_\textrm{c}} 
\newcommand{\epss}{\epsilon_\textrm{s}} 
\newcommand{\epsEX}{\epsilon_\textrm{EX}} 
\newcommand{\epsEA}{\epsilon_\textrm{EA}} 
\newcommand{\epsSA}{\epsilon_\textrm{SA}}
\newcommand{\epsest}{\epsilon_\textrm{est}}
\newcommand{\abort}{\textrm{abort}}
\newcommand{\pass}{\textrm{pass}}
\newcommand{\omegaexp}{\omega_{\textrm{exp}}}
\newcommand{\deltaest}{\delta_{\textrm{est}}}
\newcommand{\Ext}{\textrm{Ext}}
\newcommand{\Hmin}{H_{\textrm{min}}}
\newcommand{\etaopt}{\eta_\textrm{opt}}

\begin{document}
\title{Randomness extraction from Bell violation  with continuous parametric down conversion}

\author{Lijiong~Shen}
\affiliation{Centre for Quantum Technologies, National University of Singapore, 3 Science Drive 2, Singapore 117543}
\affiliation{Department of Physics, National University of Singapore, 2 Science Drive 3, Singapore 117551}

\author{Jianwei~Lee}
\affiliation{Centre for Quantum Technologies, National University of Singapore, 3 Science Drive 2, Singapore 117543}

\author{Le~Phuc~Thinh}
\affiliation{Centre for Quantum Technologies, National University of Singapore, 3 Science Drive 2, Singapore 117543}

\author{Jean-Daniel~Bancal}
\affiliation{Department of Physics, University of Basel, Klingelbergstrasse 82, 4056 Basel, Switzerland}

\author{Alessandro~Cer\`{e}}
\affiliation{Centre for Quantum Technologies, National University of Singapore, 3 Science Drive 2, Singapore 117543}

\author{Antia~Lamas-Linares}
\affiliation{Texas Advanced Computing Center, The University of Texas at Austin, Austin, Texas}
\affiliation{Centre for Quantum Technologies, National University of Singapore, 3 Science Drive 2, Singapore 117543}

\author{Adriana~Lita}
\affiliation{National Institute of Standards and Technology, Boulder 80305, CO, USA}

\author{Thomas~Gerrits}
\affiliation{National Institute of Standards and Technology, Boulder 80305, CO, USA}

\author{Sae~Woo~Nam}
\affiliation{National Institute of Standards and Technology, Boulder 80305, CO, USA}

\author{Valerio~Scarani}
\affiliation{Centre for Quantum Technologies, National University of Singapore, 3 Science Drive 2, Singapore 117543}
\affiliation{Department of Physics, National University of Singapore, 2 Science Drive 3, Singapore 117551}

\author{Christian~Kurtsiefer}
\affiliation{Centre for Quantum Technologies, National University of Singapore, 3 Science Drive 2, Singapore 117543}
\affiliation{Department of Physics, National University of Singapore, 2 Science Drive 3, Singapore 117551}

\email[]{christian.kurtsiefer@gmail.com}
\date{\today}
\begin{abstract}
We present a violation of the CHSH inequality without the fair sampling assumption with
a continuously pumped photon pair source combined with two high efficiency superconducting detectors.
Due to the continuous nature of the source, the choice of the duration of each measurement round effectively controls the average number of photon pairs participating in the Bell test.
We observe a maximum violation of
$S= 2.01602(32)$ with average number of pairs per round of $\approx 0.32$, compatible with
our system overall detection efficiencies. 
Systems that violate a Bell inequality are guaranteed to generate private randomness, with the randomness extraction rate depending on the observed violation and on the repetition rate of the Bell test.
For our realization, the optimal rate of randomness generation is a compromise
between the observed violation and the duration of each measurement round,
with the latter realistically limited by the detection time jitter.
Using an extractor composably secure against quantum adversary with quantum side information,
we calculate an asymptotic rate of $\approx 1300$ random bits/s.
With an experimental run of~$43$~minutes, we generated~617\,920~random bits, corresponding to~$\approx 240$ random bits/s.
\end{abstract}


\maketitle

Based on a violation of a Bell inequality, quantum physics can provide
randomness that can be certified to be private, i.e.,  uncorrelated to any outside process~\cite{Colbeck:2007vo,Pironio:2010bu,Acin:2016ke}. Initial experimental realizations of such sources of certified randomness are
based on atomic or atomic-like systems, but exhibit extremely low generation rates, making them impractical for most applications~\cite{Pironio:2010bu,Hensen:2015ccp}. Advances in high efficiency infrared photon detectors~\cite{Lita:2008wv, Marsili:2013fs},
combined with highly efficient photon pair sources, allowed experimental demonstrations of loophole free violation of the Bell inequality using photons~\cite{giustina2015,Shalm:2015cw}.
Due to
the small observed violation of the Bell inequality in these setups,
the random bit generation rate is on the order of tens per second in~\cite{Bierhorst:2018uk}, where they close all loopholes and are limited by the repetition rate of the polarization modulators, 
and 114~bit/s~\cite{Liu:2017uz}, where they close only the detection loophole and the main limitation is the fixed repetition rate of the photon pair source.

In this work, we use a source of polarization entangled photon pairs operating in a continuous wave (CW) mode, and define measurement rounds by organizing the detection events in uniform time bins.
The binning is set independently of the detection time, thus avoiding the coincidence loophole~\cite{Larsson:2004bt, Christensen:2015kn}.
Superconducting detectors with a high detection efficiency allow us to close the detection loophole. We show how, for fixed overall detection efficiency and pair generation rate, the time bin duration determines the observed Bell violation. We then estimate the rate of random bits that can be extracted from the system and its dependence on time bin width. 

\textit{Theory. --} Bell tests are carried out in successions of rounds. In
each round, each party chooses a measurement and records an outcome. The
simplest meaningful scenario involves two parties, each of which can choose
between two measurements with binary outcome. Alice and Bob's measurements are
labelled by $x,y\in\{0,1\}$, respectively; their outcomes are labelled $a,b\in\{+1,-1\}$. As figure of merit we use the Clauser-Horne-Shimony-Holt (CHSH) expression
\begin{align}\label{eq:bell}
    S = E_{00}+E_{01}+E_{10}-E_{11}\,,
\end{align} where the correlators are defined by
\begin{align}
    E_{xy}:=\Pr(a=b|x,y)-\Pr(a\neq b|x,y)\,.
\end{align} As well known, if $S>2$, the correlations cannot be due to pre-established agreement; and if they can't be attributed to signaling either, the underlying process is necessarily random. This is not only a qualitative statement: the amount of extractable private randomness can be quantified. In the limit in which the statistics are collected from an arbitrarily large number of rounds, the number of random bits per round, according to~\cite{Pironio:2010bu}, is at least
\begin{equation}
    r_\infty \geq 1 - \log_2\left(1+\sqrt{2-\frac{S^2}{4}}\right)\,.
\end{equation}
Tighter bounds on the extractable randomness as a function of~$S$ can be obtained by solving a sequence of semidefinite programs~\cite{Pironio:2010bu}.

Besides the no-signaling assumption, \textit{this certification of randomness is device-independent}: it relies on the value of $S$ extracted from the observed statistics, but not on any characterisation of the degrees of freedom or of the devices used in the experiment. All that matters is that in every round both parties produce an outcome. In our case, we decide that, if a party's detectors did not fire in a given round, that party will output $+1$ for that round. This convention allows us to use only one detector per party \cite{Giustina:2013uf,Bancal:2013wa}: in the rounds when the detector fires, the outcome will be~$-1$.

While the certification is device-independent, the design of the experiment requires detailed knowledge and control of the physical degrees of freedom. Our experiment uses photons entangled in polarisation, produced by spontaneous parametric down-conversion (SPDC).

Let us first consider a simplified model, in which a pair of photons is created in each round. Eberhard~\cite{Eberhard:1993tx} famously proved that, when the collection efficiencies $\eta_A$ and $\eta_B$ are not unity, higher values of $S$ are obtained using non-maximally entangled pure states. So we aim at preparing 
\begin{equation}\label{eq:eb_state} \ket{\psi} = \cos\theta \ket{HV} -
  e^{i\phi} \sin\theta\ket{VH}\,,
\end{equation}
where~$H$ and~$V$ represent the horizontal and vertical polarization modes, respectively. The state and measurement that maximise $S$ are a function of $\eta_A$ and $\eta_B$. For $\phi=0$, the optimal measurements correspond to linear polarisation directions, denoted $\cos \alpha_x \hat{e}_H + \sin \alpha_x \hat{e}_V$ and $\cos \beta_y \hat{e}_H + \sin \beta_y \hat{e}_V$.

For a down-conversion source, the number of photons produced per round is not
fixed. If the duration~$\binwidth$ of a round is much longer than the
single-photon coherence time, and no multi-photon states are generated (a
realistic assumption in a CW pumped scenario), the output of the source is accurately described by independent photon pairs, whose number~$v$ follows a Poissonian distribution $P_{\mu}(v)$ of average pairs per round $\mu$.
The main contribution to $S>2$ will come from the single-pair events; notice that $P_{\mu}(1)\leq \frac{1}{e}\approx 0.37$ for a Poissonian distribution. So there is always a large fraction of other pair number events, and the observed value of $S$ depends significantly on it~\cite{Vivoli:2015aa}. For $\mu\rightarrow 0$, almost all rounds will give no detection, that is $P(+1,+1|x,y)\approx 1$ which leads to $S=2$. So, for $\mu\ll 1$ we expect a violation $S\approx P_{\mu}(1)\,S_{\textrm{qubits}}+(1-P_{\mu}(1))2$, where $S_{\textrm{qubits}}$ is the value achievable with state \eqref{eq:eb_state}. In the other limit, $\mu\gg 1$, almost all round will have a detection, that is $P(-1,-1|x,y)\approx 1$ and again $S=2$. Before this behavior kicks in, when more than one pair is frequently present we expect a drop in the value of $S$, since the detections may be triggered by independent pairs. An accurate modelling for any value of $\mu$ is conceptually simple but notationally cumbersome; we leave it for Appendix~\ref{appmodel}.\\
Photon pair sources based on pulsing quasi-CW sources with a fixed repetition
rate control the value of $\mu$ by limiting the pump power. With true CW pumping the average number of pairs per round is
$\mu = (\textit{pair rate})\cdot \binwidth$, where $\binwidth$ is the round duration. The resulting repetition rate of the experiment is $1/\binwidth$. In this work, we fix the pair rate, while $\binwidth$ is a free parameter that can be optimized to extract the highest amount of randomness.

\begin{figure}
    \includegraphics[width=\columnwidth]{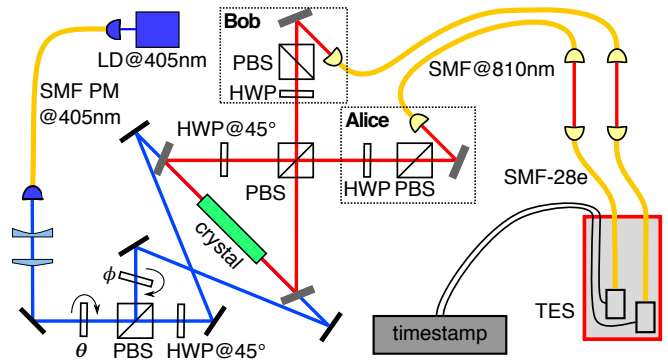}
    \caption{\label{fig:setup}
        Schematic of the experimental setup, including the source of the non-maximally entangled photon pairs.
        A PPKTP crystal, cut and poled for type II spontaneous parametric down conversion from 405~nm to 810~nm, is placed at the waist of a Sagnac-style interferometer
        and pumped from both sides.
        Light at 810~nm from the two SPDC process is overlapped in a
        polarizing beam splitter (PBS), generating the non-maximally entangled
        state described by Eq.~(\ref{eq:eb_state}) when considering a single
        photon pair.
        A laser diode (LD) provides the continuous wave UV pump light.
        The combination of a half wave plate and polarization beam splitter (PBS)
        sets~$\theta$ by controlling
        the relative intensity of the two pump beams, while
        a thin glass plate controls their relative phase~$\phi$.
        The pump beams enter the interferometer through dichroic mirrors.
        At each output of the PBS, the combination of a HWP and PBS projects the mode polarization before coupling into a fiber single mode for light at 810~nm (SMF@810).
        A free space link is used to transfer light from SMF@810 to single
        mode fibers designed for 1550~nm (SMF-28e).
        Eventually the light is detected with high efficiency superconducting Transition Edge Sensors (TES), and timestamped with a resolution of 2~ns. 
        }
\end{figure}

\textit{Experimental setup. --} A sketch of the experimental setup is shown in Fig.~\ref{fig:setup}. The source
for entangled photon pairs is based on the coherent combination of two
collinear type-II SPDC processes~\cite{Fiorentino:2004bu}.
We pump a periodically poled potassium titanylphyspate crystal (PPKTP,
$2\times1\times10$\,mm$^3$) from two opposite directions with light from the
same laser diode (405\,nm). Both pump beams have the same
Gaussian waists of~$\approx350\,\mu$m located within the crystal.
Light at 810\,nm from the two SPDC processes is overlapped in a polarizing beam
splitter (PBS), entangling the polarization modes, and collected into single
mode fibers.
When a single photon pair is generated, the resulting polarization state is given by
Eq.~\eqref{eq:eb_state}, where $\theta$ and~$\phi$ are determined by the
relative intensity and phase of the two pump beams set by rotating a half wave
plate before the first PBS, and the tilt of a glass plate in one of the pump
arms.

The effective collection modes for the downconverted light, determined by the
single mode optical fibers and incoupling optics was chosen to have a Gaussian
beam waist of $\approx130~\mu$m centered in the crystal in order to maximize
collection efficiency~\cite{Bennink:2010kd,Dixon:2014bg}.
The combination of a zero-order half-wave plate and another PBS (extinction
rate 1:1000 in transmission
) sets the measurement bases
for light entering the single mode fibers.
All optical elements are anti-reflection coated for 810\,nm.
Light from each collection fiber is sent to a superconducting transition edge
sensor (TES) optimized for detection at 810\,nm~\cite{Lita:2008wv}, which are
kept at $\approx80$~mK within a cryostat.
As the detectors show the highest efficiency when coupled to telecom fibers
(SMF28+), the light collected in to single mode fibers from the parametric
conversion source is transferred to these fibers via a free-space link.
The TES output signal is translated into photodetection event arrival times
using a constant fraction discriminator with an overall timing jitter $\approx 170$\,ns,
and recorded with a resolution of 2\,ns.
Setting Alice's and Bob's analyzing waveplates in the natural basis of the
combining PBS, $HV$ and $VH$, we estimate heralded efficiencies of
$82.42\pm0.31$~\% ($HV$) and $82.24\pm0.30$~\% ($VH$).
We identified two main sources of uncorrelated detection events: intrinsic
detector and background events at rates of $6.7\pm0.58\,{\text{s}}^{-1}$ for Alice and $11.9\pm0.77\,{\text{s}}^{-1}$
for Bob, respectively, and fluorescence caused by the UV pump in
the PPKTP crystal~\cite{Hegde:2007jc}, contributing $0.135\pm 0.08\%$ of the signal.
With a total pump power at the crystal of 5.8\,mW we estimate a pair
generation rate~$\approx2.4\times 10^4\,{\text{s}}^{-1}$ (detected $\approx20\times 10^3\,{\text{s}}^{-1}$),
and dark count / background rates of 45.7\,${\text{s}}^{-1}$ (Alice) and
41.5\,${\text{s}}^{-1}$ (Bob).
\begin{figure}
    \includegraphics[width=1\columnwidth]{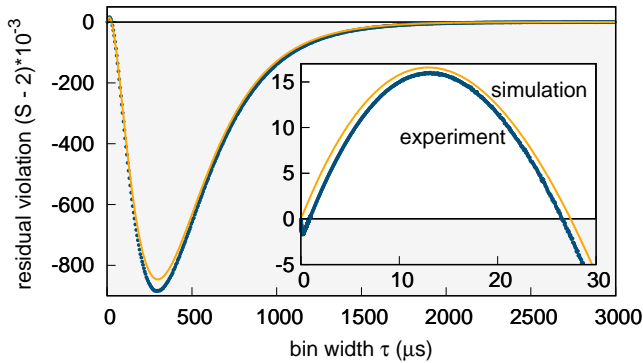}
    \caption{\label{fig:vio_bin}
        Measured CHSH violation as function of bin width $\binwidth$ (blue circles).
        A theoretical model (orange continuous line) is sketched in the main
        text and described in detail in Appendix \ref{appmodel}.
        Both the simulation and the experimental data show a violation for
        short $\binwidth$ (zoom in inset).
        The uncertainty on the measured value, calculated assuming i.i.d., corresponding
        to one standard deviation due to a Poissonian distribution of the
        events, is smaller than the symbols. For~$\binwidth\lesssim 1~\mu$s the detection jitter ($\approx 170$~ns) is comparable with the time bin, resulting in a loss of observable correlation and a fast drop of the value of $S$.}
\end{figure}

\textit{Violation. --} For the measured system efficiencies ($\eta_A\approx 82.4\%$, $\eta_B\approx 82.2\%$)
and rate of uncorrelated counts at each detector
(45.7\,${\text{s}}^{-1}$ Alice, 41.5\,${\text{s}}^{-1}$ Bob),
a numerical optimisation gives the following values of the state and
measurement parameters (see Appendix \ref{appmodel} for details):
$\theta = 25.9^{\circ}$, $\alpha_0 = -7.2^{\circ}$,
$\alpha_1 = 28.7^{\circ}$,
$\beta_0 = 82.7^{\circ}$, and
$\beta_1 = -61.5^{\circ}$. These are close to optimal for all values of~$\mu$, and the maximal violation is expected for~$\mu= 0.322$. 

We collected data for approximately~$42.8$~minutes, changing the measurement basis every 2~minutes,
cycling through the four possible basis combinations.
The sequence of the four settings is determined for every cycle using a pseudo-random number generator. 
We periodically ensure that~$\phi\approx 0$ by rotating the phase plate until the visibility in the~$+45^\circ/-45^\circ$ basis is larger than~0.985. Excluding the phase lock, the effective data acquisition time is $\approx34$~min.

In Fig.~\ref{fig:vio_bin} we show the result of processing the timestamped events for different bin widths~$\binwidth$. The largest violation $S=2.01602(32)$ is observed for~$\binwidth = 13.150~\mu$s, which, with the cited pair generation rate of $24\times 10^3\,$s$^{-1}$, corresponds to $\mu\approx 0.32$.
The uncertainty is calculated assuming that measurement results are independent and identically distributed (i.i.d.). Since the fluctuations of $S$ are identical in the i.i.d. and non-i.i.d. settings, this uncertainty is also representative of the p-value associated with local models~\cite{Bierhorst:2015fq,Elkouss:2016ii}.
The slight discrepancy between the experimental violation and the simulation is attributed to the non-ideal visibility of the state generated by the photon pair source. When $\binwidth$ is comparable to the detection jitter, detection events due to a single pair may be assigned to different rounds, decreasing the correlations. This explains the drop of $S$ below 2 (which our simulation does not capture because we have not included the jitter as a parameter).

\begin{figure}
    \includegraphics[width=1\columnwidth]{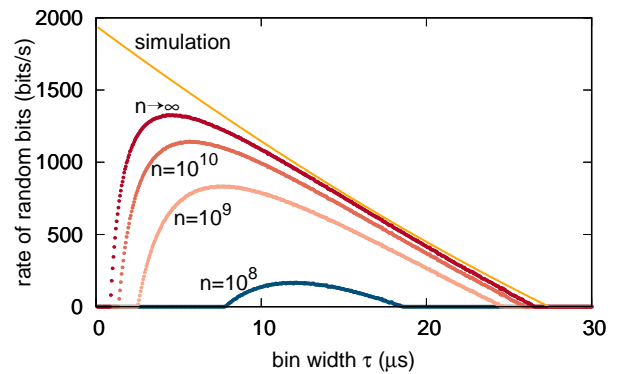}
    \caption{\label{fig:rnd_bin}
        Randomness generation rate~$r_n/\binwidth$ as a function of~$\binwidth$
        for different block sizes~$n$.
        The points are calculated via Eq.~\eqref{eq:rand_finite} for finite
        $n$ (Eq.~\eqref{eq:rand_asymp} for $n\to\infty$) and the 
        violation measured in the experiment,
        assuming $\gamma=0$ (no testing rounds) and
        $\epsc=\epss=10^{-10}$.
        The continuous line is the asymptotic rate Eq.~\eqref{eq:rand_asymp}
        evaluated on the values of~$S$ of the simulation shown in Fig.~\ref{fig:vio_bin}, for the same security assumptions.
        }
\end{figure}

\textit{Randomness extraction. --} In order to turn the output data generated
from our experiment into uniformly random bits, we need to employ a randomness
expansion protocol~\cite{Rotem:diproofs}. Such a protocol consists of a pre-defined number of rounds~$n$, forming a block. Each round is randomly assigned (with probability $\gamma$ and $1-\gamma$, respectively) to one of two tasks: testing the device for faults or eavesdropping attempts, or generating random bits. When the test rounds show a sufficient violation, one applies a quantum-proof randomness extractor to the block, obtaining~$m$ random bits. 
The performance of the extraction protocol is determined by completeness and soundness security parameters, $\epsc$ and $\epss$. To ensure the resulting string is uniform to within $\approx10^{-10}$, we choose $\epsc=\epss=10^{-10}$.
The extraction protocol is a one-shot extraction protocol, i.e., the security analysis does not assume i.i.d..
The output randomness is composable and secure against a quantum adversary holding quantum side information~\cite{Rotem:diproofs}. The details of the protocol execution and its security proof are given in Appendix~\ref{sec:proofs}.

For a block consisting of $n$ rounds, the number of random bits per round is at least
\begin{align}
    \label{eq:rand_finite}
    r_n = \eta_\textrm{opt}(\epsilon',\epsEA) -4\frac{\log n}{n} + 4\frac{\log\epsEX}{n} -\frac{10}{n}\,,
\end{align} 
where the function $\etaopt$ depends on the block size~$n$, detected
violation~$S$, and auxiliary security parameters~$\epsilon'$, $\epsEA$, $\epsEX$. The choice of these auxiliary security parameters
is required to add up to the chosen level of completeness and soundness.
In the limit~$n\to\infty$ we obtain a lower bound on the number of random bits per round
\begin{equation}\label{eq:rand_asymp}
    r_\infty =
    \;1-h\left(\frac{1}{2}+\frac{1}{2}\sqrt{\frac{S^2}{4}-1}\right)\,,
\end{equation}
where $h(p) := - p\,\log_2p - (1 - p)\log_2(1 - p)$ is the binary entropy function. 

The extractable randomness rate~$r_n/\binwidth$ based on the observed~$S$ is presented
in Fig.~\ref{fig:rnd_bin} for various block sizes~$n$. For comparison, we also
plot the asymptotic value~$r_\infty/\binwidth$ with~$S$ given by the simulation.
The most obvious feature is that the highest randomness rate is not obtained at maximal violation of the inequality. There one gets highest randomness
per round, but it turns out to be advantageous to sacrifice randomness per round in favor of a larger number of rounds per unit time. This optimization will be part of the calibration procedure for a random number generator with an active switch of measurement bases.
As explained previously, the detection jitter affects the observable violation for~$\binwidth$ comparable to it.
This causes the sharp drop for short time bins observed for the experimental data.
For fixed detector efficiencies,
we expect the randomness rate to increase with higher photon pair generation
rate, that is by increasing the pump power, and to be ultimately limited by
the detection time jitter. Here, the use of efficient superconducting nanowire detectors will be a significant advantage.

We generated a random string from the data used to demonstrate the violation.
We sacrificed~$\approx22$\% of the data as calibration to determine the optimal bin width ($8.9~\mu$s),
and estimate the corresponding violation.
We
applied the extractor to the remaining~$\approx78$\% of the data, corresponding to 
175\,288\,156~bins, 
obtaining 617\,920~random bits.
From the total measurement time of 42.8~min,
we calculate a rate of~$\approx240$ random bit/s.
For details of the extraction process see Appendix~\ref{appextraction}.
Considering only the net measurement time,
that is without
the acquisition of the calibration fraction of the data, the phase lock of the source, and the rotation of waveplate motors,
we obtain a randomness rate of~$\approx396$~bit/s.
These numbers are not necessarily optimal; more sophisticated analysis demonstrated randomness extraction for very low detected violations~\cite{Bierhorst:2018uk, martingale_randomness}, and may yield a larger extractable randomness also in our case.
Details of the extraction procedure are in Appendix~\ref{appextraction}.

\textit{Conclusion. --}
We experimentally observed a violation of CHSH inequality with a continuous wave photon entangled pair source without the fair-sampling assumption combining a high collection efficiency source and high detection efficiency superconducting detectors,
with the largest detected violation of~$S=2.01602(32)$.

The generation rate
of all probabilistic sources of entangled photon pairs is limited by the probability of generation of multiple pairs
per experimental round, according to Poissonian statistics.
The flexible definition of an experimental round permitted by the CW nature of our setup allowed us to study the dependence of the observable violation as function of the average number of photon pairs per experimental round.
This same flexibility can be exploited to reduce the time necessary to acquire sufficient statistics for this kind of experiments: an increase in the pair generation rate is accompanied by a reduction of the
round duration.
This approach shifts the experimental repetition rate limitation from the photon statistics to the other elements of the setup, e.g. detectors time response or active polarization basis switching speed.

The observation of a Bell violation also certifies the generation of randomness.
We estimate the amount of randomness generated per round both in an asymptotic
regime and for a finite number of experimental rounds, assuming a required level of uniformity of $10^{-10}$.
When considering the largest attainable \textit{rate} of random bit generation,
the optimal round duration is the result of 
the trade-off between observed violation and number of rounds per unit time.
While for an ideal realization the optimal
round duration would be infinitesimally short,
it is limited in our system by the detection jitter time.
Our proof of principle demonstration can be extended into a complete, loophole-free random number source.
This requires closing the locality and freedom-of-choice loopholes, with techniques not different from pulsed photonic-sources, with the only addition of a periodic calibration necessary for determining the optimal time-bin.

\section*{Acknowledgments} This research is supported by the Singapore Ministry of Education Academic Research Fund Tier 3 (Grant No.~MOE2012-T3-1-009); by the National Research Fund and the Ministry of Education, Singapore, under the Research Centres of Excellence programme; by the Swiss National Science Foundation (SNSF), through the Grants PP00P2-150579 and PP00P2-179109; and by the Army Research Laboratory Center for Distributed Quantum Information via the project SciNet.
This work includes contributions of the National Institute of Standards and Technology, which are not subject to U.S. copyright.

\bibliographystyle{apsrev4-1}
%

\appendix

\section{Modelling the violation of CHSH by a Poissonian source of qubit pairs}
\label{appmodel}

The output of a CW-pumped SPDC process can be accurately described as the emission of independent pairs distributed according to Poissonian statistics of average $\mu$, if the time under consideration (in our case, the length of a round) is much longer than the single-photon coherence time. The goal of this appendix is to provide an estimate of the observed CHSH parameter $S$ for such a source.

The pairs being independent, it helps to think in two steps. First, each pair is converted into classical information $(\alpha,\beta)\in\{+,-\}$ with probability
\begin{align}
    P_Q(\alpha,\beta|x,y)=\textrm{Tr}(\rho \Pi_\alpha^x\otimes \Pi_\beta^y)\,,
\end{align}
where the $\Pi$'s are measurement operators. If some of the events have $\alpha=-$ ($\beta=-$), Alice's (Bob's) detector may be triggered, leading to the observed outcome $a=-1$ ($b=-1$).

For the purpose of studying CHSH, it is sufficient to consider $P(-1,+1|x,y)$ and $P(+1,-1|x,y)$, since
\begin{align}
    E_{xy}=1-P(-1,+1|x,y)-P(+1,-1|x,y)\,.
\end{align}
With our convention of outcomes, $P(-1,+1|x,y)$ is the
probability associated with the
case when Alice's
detector clicks and Bob's does not.
Thus, Bob's detector should not be triggered by any pair: each pair will contribute to $P(-1,+1|x,y)$ with
\begin{align}
    D(\alpha)\equiv P_Q(\alpha,+|x,y)+(1-\eta_B)P_Q(\alpha,-|x,y)\,.
\end{align}
Now, let us look at the contribution by $v$ pairs to $P(-1,+1|x,y)$. At least one of the $\alpha$'s must be $-$ for the detector to be triggered; and multiple detections will also be treated as $a=-1$. Thus, a configuration in which exactly $k$ $\alpha$'s are $-$ leads to $a=-1$ with probability $1-(1-\eta_A)^k$ (i.e. at least one $\alpha=-$ must trigger a detection). Obviously there can be $\binom{v}{k}$ such configurations, so the contribution of the $v$ pair events to $P(-1,+1|x,y)$ is
\begin{equation}\label{eq:Dn}
    D_v=\sum_{k=1}^v \binom{v}{k} [1-(1-\eta_A)^k] D(-)^kD(+)^{v-k}\,.
\end{equation}
Finally,
\begin{equation}
    P(-1,+1|x,y)=\sum_{v=0}^{\infty} P_{\mu}(v)D_v\,.
\end{equation} The calculation of $P(+1,-1|x,y)$ is identical, with $D(\beta)\equiv P_Q(+,\beta|x,y)+(1-\eta_A)P_Q(-,\beta|x,y)$ and $\eta_B$ instead of $\eta_A$ in \eqref{eq:Dn}.

Because the quantum probabilities appear in such a convoluted way, the optimal parameters for both the state and the measurements are not the same as for the single-pair case. Upon inspection, however, the values are close, as expected from the fact that the violation is mostly contributed by the single-pair events.

The curves presented in Fig.~\ref{fig:vio_bin} have been obtained with a slightly modified model that
includes the effect of the background events. The quantum probabilities $P_Q(\alpha,\beta|x,y)$ have been computed with $\rho=\ket{\psi}\bra{\psi}$ given in \eqref{eq:eb_state} and with projective measurements, with the values of the parameters given in the main text.

\section{\label{sec:proofs}Protocol and security proof}

For completeness, we will present the protocol studied in~\cite{Rotem:diproofs} and give explicit constants in its security proof. We also refer to this paper for basic definitions of (smooth) min-entropy and related quantities. It will be more convenient for us to switch to to the notation $a,b\in\{0,1\}$ for the outcome labels (instead of $a,b\in\{+1,-1\}$ of the main text) and use the language of nonlocal games with winning condition
\begin{align}
     w_\textrm{CHSH}(a,b,x,y) = \begin{cases} 1 &\textrm{if } a\oplus b=x\cdot y\,,\\
     0 &\textrm{otherwise}\,.
     \end{cases}
\end{align}
The game winning probability is then $w=1/2+S/8$ in terms of the CHSH value. The optimal classical winning strategy achieve a winning probability of 0.75, while the optimal quantum strategy achieves a winning probability of $(2+\sqrt{2})/4\approx0.85$.

A randomness expansion protocol is a procedure that consumes $r$-bits of randomness and generates $m$-bits of almost uniform randomness. Formally, a $(\epsc,\epss)$-secure $r\to m$ randomness expansion protocol if given $r$ uniformly random bits,
\begin{itemize}
    \item (Soundness) For any implementation of the device it either aborts or returns an $m$-bit string $Z\in\{0,1\}^m$ with
    \begin{align*}
        (1-\Pr[\abort])\norm{\rho_{ZRE}-\rho_{U_m}\otimes\rho_{U_r}\otimes\rho_E}_1  \leq \epss\,,
    \end{align*}
    where $R$ is the input randomness register, $E$ is the adversary system, and $\rho_{U_m}, \rho_{U_r}$ are the completely mixed states on appropriate registers.
    \item (Completeness) There exists an honest implementation with $\Pr[\abort]\leq\epsc$.
\end{itemize}
We remark that this security definition is a composable definition assuming quantum adversary, but not composable assuming a no-signalling adversary~\cite{Rotem:diproofs}. Composability allows the randomness generated to be safely used inside a larger protocol, such as quantum key distribution, without compromising the latter's security.

For a concrete randomness expansion protocol, we present the protocol studied in~\cite{Rotem:diproofs}. The protocol takes parameters $\gamma$ expected fraction (marginal probability) of test rounds, $\omegaexp$ expected winning probability for an honest (perhaps noisy) implementation, and $\deltaest$ width of the statistical confidence interval for the estimation test. In an execution, for every round $i\in\{1,\dots,n\}$:
\begin{itemize}
    \item Bob chooses a random bit $T_i\in\{0,1\}$ such that $\Pr(T_i=1)=\gamma$ using the interval algorithm~\cite{interval_algorithm}.
    \item If $T_i=0$ (randomness generation), Alice and Bob choose deterministically $(X_i,Y_i)=(0,0)$, otherwise $T_i=1$ (test round) they choose uniformly random inputs $(X_i,Y_i)$.
    \item Alice and Bob use the physical devices with the said inputs $(X_i,Y_i)$ and record their outputs $(A_i,B_i)$.
    \item If $T_i=1$, they compute 
    \begin{align}
        C_i=w_\textrm{CHSH}(A_i,B_i,X_i,Y_i)\,.
    \end{align}
\end{itemize}
They abort the protocol if $\sum_jC_j<(\omegaexp\gamma-\deltaest)n$ where $j$ is the index of test rounds, otherwise they return $\Ext({\bf AB},{\bf Z})$ where $\textrm{Ext}$ is a randomness extractor, ${\bf AB}=A_1B_1...A_nB_n$ and  ${\bf Z}$ is a uniformly random seed.

More precisely, we use a Trevisan extractor in~\cite{Mauerer:extractors} based on polynomial hashing with block weak design, because of its efficiency in terms of seed length. This is a function $\Ext:\{0,1\}^{2n}\times\{0,1\}^{d}\to\{0,1\}^{m}$ such that if $\Hmin({\bf AB}|E)\geq 4\log\frac{1}{\epsilon_1}+6+m$ then
\begin{align}
    \frac{1}{2}\norm{\rho_{\Ext({\bf AB},{\bf Z}){\bf Z}E}-\rho_{U_m}\otimes\rho_{U_d}\otimes\rho_E}_1 \leq m\epsilon_1
\end{align}
The seed length of this extractor is $d = a(2\ell)^2$ where
\begin{align}
    a &= \ceil*{\frac{\log(m-2e)-\log(2\ell-2e)}{\log(2e)-\log(2e-1)}}\\
    \ell &= \ceil*{\log 2n+2\log\frac{2}{\epsilon_1}}
\end{align}

Now it can be shown that the entropy accumulation protocol gives the completeness and soundness of our randomness expansion protocol. However, let us mention how the input randomness affects the soundness and completeness of the final protocol.

In the protocol we assume access to a certain uniform randomness source, from which the random bits required in the protocol are generated: the $T_i$, $X_i$ and $Y_i$. In certain rounds, $X_i$ and $Y_i$ are either deterministic or fully random bits and can be directly obtained from the source. On the other hand, $T_i$ must be simulated from the uniform source (except when $\gamma=1/2$ which is not usually the case in practice). This can be done efficiently by the interval algorithm~\cite{interval_algorithm}: the expected number of random bits needed to generate one Bernoulli$(\gamma)$ is at most $h(\gamma)+2$ and the maximum number of random bits needed is at most $L_{\max}:=\max\{\log\gamma^{-1},\log(1-\gamma)^{-1}\}$. Then Lemma 16 of~\cite{Rotem:diproofs} gives us: let $\gamma>0$, for any $n$ there is an efficient procedure that given (at most) $6h(\gamma)n$ uniformly random bits either it aborts with probability at most $\epsSA = \exp(-18h(\gamma)^3n/L_{\max})$ or outputs $n$ bits $T_1,\dots,T_n$ whose distribution is within statistical distance at most $\epsSA$ of $n$ i.i.d. Bernoulli$(\gamma)$ random variables. This raises both the completeness and soundness parameter of the final protocol by $\epsSA$.

In an honest implementation of the protocol, Alice and Bob execute the protocol with a device that performs i.i.d. measurements on a tensor product state resulting in an expected winning probability $\omegaexp$. Here Lemma 8 of~\cite{Rotem:diproofs} bounds the probability of aborting using Hoeffding's inequality. That is, the probability that our randomness expansion protocol aborts for an honest implementation is
\begin{align}
    \Pr[\abort] \leq \exp(-2n\delta_{\textrm{est}}^2)=:\epsest\,.
\end{align}
Therefore, the total completeness is bounded by $\epsSA+\epsest$. (Note that $\epsest$ is actually the completeness parameter of the entropy accumulation protocol in~\cite{Rotem:diproofs}.)

For the soundness, Corollary 11 of~\cite{Rotem:diproofs} ensures that for any $\epsEA,\epsilon'\in(0,1)$ either the protocol aborts with probability greater than $1-\epsEA$ or
\begin{align}
    \Hmin^{\epsilon'}({\bf AB}|{\bf XYT}E)_{\rho_{|\pass}} > n\cdot\eta_\textrm{opt}(\epsilon',\epsEA)\,.
\end{align}
Together with our extractor, for all $\epsilon_1\in(0,1)$, if the length $m$ of the final string satisfies
\begin{align}
    n\cdot\eta_\textrm{opt}(\epsilon',\epsEA) = 4\log\frac{1}{\epsilon_1}+6+m
    \label{eq:rand_length_relation}
\end{align}
then we are guaranteed that
\begin{align}
    \frac{1}{2}\norm{\rho_{SRE}-\rho_{U_m}\otimes\rho_{U_R}\otimes\rho_E}_1 \leq \epsilon'/2 + m\epsilon_1\,.
\end{align}
Here $\eta_\textrm{opt}(\epsilon',\epsEA)$ is given by the following equations: for $h$ the binary entropy and $\gamma,p(1)\in(0,1]$
\begin{widetext}
\begin{align}
    \eta_\textrm{opt}(\epsilon',\epsEA) &= \max_{\frac{3}{4}<\frac{p_t(1)}{\gamma}<\frac{2+\sqrt{2}}{4}} \eta(\omega_{\textrm{exp}}\gamma-\delta_{\textrm{est}},p_t,\epsilon',\epsEA)\,,\\
    \eta(p,p_t,\epsilon',\epsEA) &= f_{\min}(p,p_t)-\frac{1}{\sqrt{n}}2\left(\log13+\frac{\diff}{\diff p(1)}g(p)|_{p_t}\right)\sqrt{1-2\log(\epsilon'\epsEA)}\,,\\
    f_{\min}(p,p_t) &= \begin{cases}
    g(p) &\textrm{ if } p(1)\leq p_t(1)\,,\\
    \frac{\diff}{\diff p(1)}g(p)|_{p_t}\cdot p(1) + \left(g(p_t)-\frac{\diff}{\diff p(1)}g(p)|_{p_t}\cdot p_t(1)\right) &\textrm{ if } p(1)>p_t(1)
    \end{cases}\\
    g(p) &= \begin{cases}
    1-h\left(\frac{1}{2}+\frac{1}{2}\sqrt{16\frac{p(1)}{\gamma}\left(\frac{p(1)}{\gamma}-1\right)+3}\right) &\textrm{ if } \frac{p(1)}{\gamma} \in \left[0,\frac{2+\sqrt{2}}{4}\right] \\
    1  &\textrm{ if } \frac{p(1)}{\gamma} \in \left[\frac{2+\sqrt{2}}{4},1\right]
    \end{cases}
\end{align}
\end{widetext}
Combined with the input sampling soundness, the total soundness is bounded by $\epsSA+\epsEA+\epsilon'/2 + m\epsilon_1$.

\begin{table}[h]
    \centering
    \begin{tabular}{|c|l|}
        \hline
        Parameter & \multicolumn{1}{c|}{Definition}\\
        \hline
        $\epsc$ & completeness, bounding honest abort probability\\
        $\epss$ & soundness, bounding randomness security\\
        $\epsSA$ & input sampling error tolerance\\
        $\epsilon'$ & smoothing parameter\\
        $\epsilon_1$ & 1-bit extractor error tolerance\\
        $\epsEX$ & randomness extractor error tolerance\\
        $\epsest$ & Bell estimation error tolerance\\
        $\epsEA$ & soundness of entropy accumulation protocol\\ 
        \hline
    \end{tabular}
    \caption{Definition of security parameters.}
    \label{tab:security_params}
\end{table}

Finally, let us count the number of random bits consumed in the protocol. It consists of the randomness used to decide if a round is a test or generation round, the randomness used to pick the inputs in a test round, and the randomness used for the Trevisan extractor. Taking into account the finite statistical fluctuations, we need at most $6h(\gamma)n$ bits to choose between test and generation except with probability $\epsSA$. This results in at most $2\gamma n$ testing rounds except with probability $\epsSA$, which equates to $2\times 2\gamma n$ random bits being consumed for generating the inputs for test rounds. The randomness for Trevisan extractor is $d$ bits. (Practically, after the first run of the protocol, we can omit this amount because the extractor is a strong extractor: we can reuse the seed for next run of the protocol). Summing these up, we have consumed at most $6h(\gamma)n+4\gamma n+d$ uniformly random bits with probability at least $1-2\epsSA$.

In summary, for a device with $\omegaexp$, any choice of $\gamma,\epsilon_1,\epsilon',\epsEA\in(0,1)$, and $n$ large enough, our protocol is an $(\epsSA+\epsest,\epsSA+\epsEA+\epsilon'/2 + m\epsilon_1)$-secure $[6h(\gamma)n+4\gamma n+d]\to m$ randomness expansion protocol. That is either our protocol abort with probability greater than $1-\epsEA$, or it produces a string of length $m$ such that $\frac{1}{2}\norm{\rho_{SRE}-\rho_{U_m}\otimes\rho_{U_R}\otimes\rho_E}\leq\epsSA+\epsilon'/2 + m\epsilon_1$\,. The protocol consume at most $6h(\gamma)n+4\gamma n+d$ uniformly random bits with probability at least $1-2\epsSA$.\\

\section{\label{sec:randomness_formula}Input/Output randomness analysis}
The previous Appendix gives a complete picture of the (one-shot) behavior of
our randomness expansion protocol. For the purpose of this paper, it suffices
to obtain rough estimates on the randomness output, but further optimization
can be done.

For simplicity, we introduce some bounds on the resources. Since $m\leq2n$ we can let $m\epsilon_1\leq2n\epsilon_1=:\epsEX$ which gives $\epsilon_1=\epsEX/(2n)$. Plugging this back in~\eqref{eq:rand_length_relation} gives us the number of random bits one can extract,
\begin{align}
    m = n\cdot\eta_\textrm{opt}(\epsilon',\epsEA) -4\log n + 4\log\epsEX -10\,,
    \label{eq:rand_length}
\end{align}
for a given level of soundness $\epsSA+\epsEA+\epsilon'/2+\epsEX$. Moreover, the protocol consumes $6h(\gamma)n+4\gamma n + d$  bits of randomness with probability at least $1-\epsSA$, where
\begin{align}
    d = a(2\ell)^2 &\textrm{ with } a \leq \frac{\log(2n-2e)-\log(2\ell-2e)}{\log(2e)-\log(2e-1)} + 1 \nonumber \\
    &\textrm{ and } \, \, \, \ell \leq 3\log n +6 - 2\log\epsEX\,.
    \label{eq:rand_usage}
\end{align}
This leads to an expansion of $m-6h(\gamma)n-4\gamma n - d$. Hence, the output randomness rate per unit time is
\begin{align}
    r_n = \frac{1}{\binwidth}\left(\eta_\textrm{opt}(\epsilon',\epsEA) -4\frac{\log n}{n} + 4\frac{\log\epsEX}{n} -\frac{10}{n}\right)\,,
\end{align}
and the net randomness rate per unit time is
\begin{widetext}
\begin{align}
    r^\textrm{net}_n = \frac{1}{\binwidth}\left(\eta_\textrm{opt}(\epsilon',\epsEA) -4\frac{\log n}{n} + 4\frac{\log\epsEX}{n} -\frac{10}{n} - 6h(\gamma) - 4\gamma - \frac{d}{n}\right)  \,.
\end{align}
\end{widetext}
These formulas are of course given for a protocol with completeness $\epsSA+\epsest$ and soundness $\epsSA+\epsEA+\epsilon'/2+\epsEX$, where
\begin{align}
    \epsSA &= \exp(-18h(\gamma)^3n/L_{\max})\label{eq:epsSA}\\
    L_{\max}&=\max\{\log\gamma^{-1},\log(1-\gamma)^{-1}\}\\
    \epsest &= \exp(-2n\delta_{\textrm{est}}^2)\label{eq:deltaest}\,.
\end{align}

The asymptotic rate for the block size $n\to\infty$ is given by taking the limit of block size $n$
\begin{equation}
    r_\infty = \frac{1}{\binwidth} \;\left[1-h\left(\frac{1}{2}+\frac{1}{2}\sqrt{\frac{S^2}{4}-1}\right)\right]\,.
\end{equation}
For the net asymptotic rate we could also take the same limit, however we could obtain a better bound by the expected behavior of the interval algorithm. Since the expected number of random bits needed to generate $T_1,...,T_n$ is $nh(\gamma)+2$ by~\cite{interval_algorithm}, and $\gamma n$ of which is expected to be test rounds each consuming 2 random bits, we have the asymptotic net rate
\begin{equation}
    r_\infty^\textrm{net} = \frac{1}{\binwidth} \;\left[1-h\left(\frac{1}{2}+\frac{1}{2}\sqrt{\frac{S^2}{4}-1}\right)-h(\gamma) - 2\gamma\right]\,.
\end{equation}

From an end-user perspective, one may argue that the only parameters of
interest are the completeness and soundness security parameters which will
constrain the rest of protocol parameters---$\gamma,\deltaest,n,\epsilon$'s---for a given objective such as maximizing randomness rate or net randomness rate. For the illustrative plots we set the
constraints 
\begin{align}\label{eq:epsilons}
    \epsSA+\epsEA+\epsilon'/2+\epsEX &= \epss\,,\\
    \epsSA+\epsest &= \epsc\,,
\end{align}
and fix $\epsc=10^{-10},\epss=10^{-10}$.
One then maximizes randomness output or net randomness output given these constraints. This gives us the best (as measured by our objective function) protocol parameters within the relaxations made to obtain~\eqref{eq:rand_length} and~\eqref{eq:rand_usage}.

However, in the main text we take a simpler approach without optimizing over the variables $\gamma,\deltaest,\epsilon$'s. For each block size $n$, we compute $\epsSA$ as given by~\eqref{eq:epsSA} which then fixes $\deltaest$ via $\epsest=10^{-10}-\epsSA$ and~\eqref{eq:deltaest}. The remaining $\epsilon$'s which have weight $10^{-10}-\epsSA$ are chosen in a $1:2:1$ ratio of $\epsEA:\epsilon':\epsEX$, which is guaranteed to add up to the specified level of completeness and soundness. This approach is not far from optimal in the regime of large block size $n$. This results in the
experimental points reported in Figure~\ref{fig:rnd_bin}.

\section{Random bits extraction procedure}
\label{appextraction}

As mentioned in the main text, the data observed during the experiment contains certified randomness. Here we describe the procedure we use to extract this randomness in a finite run of the experiment. We consider two blocks of data, corresponding to an acquisition time of $\approx42.8$ min (\texttt{dataset1}) and $\approx17.33$ hours (\texttt{dataset2}).

The randomness protocol we use (described in Appendix B) relies on two elements:
\begin{itemize}
\item an honest implementation, and
\item security parameters.
\end{itemize}
These elements must be chosen adequately before proceeding to the extraction. Indeed, if a too optimistic honest implementation is chosen, for instance, the data observed will fail to pass the test, and the whole protocol will abort: no randomness can then be extracted.

Moreover, our setup allows us to choose freely the
\begin{itemize}
\item bin width
\end{itemize}
which can significantly affect the amount of certified randomness.

We dedicate a fraction $\gamma_\text{calib}$ of our data to the estimation of these parameters
so as to maximize the amount of randomness certified.
The randomness protocol is then run with these parameters on the remaining fraction $(1-\gamma_\text{calib})$ of the data only.
We determine
the fraction of data~$\gamma_\text{calib}$ to use for the calibration of the randomness extraction procedure
from a simulation of the experiment.
We estimate the number of random bits that can be certified from an experiment of the envisioned length if a fraction $\gamma_\text{calib}$ of the data is dedicated to calibration purpose (all other parameters being set as expected). We choose the value of $\gamma_\text{calib}$ that maximizes this quantity. We find that $\gamma_\text{calib}=22\%$ is adequate for
\texttt{dataset1}, and $\gamma_\text{calib}=8\%$ for \texttt{dataset2}.

We then proceed to define the parameters of the randomness protocol.
The security parameters $\epss, \epsc$ are set a priori, with all the other parameters derived as described in Appendices~\ref{sec:proofs} and~\ref{sec:randomness_formula}, with~$\gamma=1$.
We define \textit{honest implementation}
an implementation which reproduces the CHSH violation observed during the calibration stage with probability
\begin{equation}
P(S_\text{exp}\geq S_\text{calib})\geq \epsilon_\text{calib}\,.
\end{equation}
For concreteness, we set $\epsilon_\text{calib}=10^{-10}$. This step guarantees that we will not overestimate the amount of Bell violation which we can expect from a honest experiment. This is crucial for the whole certification procedure to succeed with a large probability. We then have
\begin{equation}
w_\text{exp} = w_\text{calib} - \delta_\text{calib}\,,
\end{equation}
with $\delta_\text{calib} = B(\omega_\text{exp},(1-\gamma_\text{calib})n,\omega_\text{calib})$, where
\begin{equation}
B(p,n,q) = \sum_{i=0}^{nq}\binom{n}{i}p^i(1-p)^{n-i}
\end{equation}
is the cumulative distribution of $n$ Bernoulli variable with parameter $p$. For simplicity, we use the upper bound
\begin{equation}
\delta_\text{calib}\leq \sqrt{\frac{\log(1/\epsilon_\text{calib})}{2n}}
\end{equation}
valid for all winning probability~$\omega_\text{calib}$, which leads to a conservative estimate of the honest implementation Bell violation~$S_\text{est}$.

Having fixed all security parameters and defined our honest implementation, we are now left with the choice of the bin width. For this, following the procedure discussed in the main text, we compute the number of certified random bits that we can hope to certify in the remaining $(1-\gamma_\text{calib})$ fraction of the data as a function of the bin width. We then choose the bin width which yields the maximum rate of random bits.
We find that optimal bin widths 8.9~$\mu$s for \texttt{dataset1} and 5.35~$\mu$s for \texttt{dataset2}.
This allows us to define how the remaining data is to be treated: first, we extract the outcomes corresponding to the chosen bin width, then we use the exact number of bins so extracted to compute precisely the threshold Bell violation $w_\text{exp}-\delta_\text{est}$ and the number of certified bits $m$ corresponding to this dataset, finally we check whether the data indeed yields a Bell violation larger than $w_\text{exp}-\delta_\text{est}$.
If this is not the case, we abort. Otherwise, we apply a randomness extractor on the string of outcomes.
Both datasets pass this test.

Finally, we use the Trevisan extractor implemented by Mauerer et al.~\cite{Mauerer:extractors}, and further improved by Bierhorst et al.~\cite{Bierhorst:2018uk} to extract the certified bits.
The advantage of Trevisan extractors over other kinds of randomness extractors is that they require little initial seeds, and that they are composable, strong extractors, and secure against quantum side information. Following the suggestion of~\cite{Mauerer:extractors}, we use the block weak design construction with a RSH extractor to maximize the number of extracted bits. The extractor also require a supply of seed randomness. For this, we use the bits generated with the random number generator described in~\cite{Shi:2016kw}.

In the end, we extract 617\,920 and 35\,799\,872 uniformly random bits from \texttt{dataset1} and \texttt{dataset2} respectively, using seeds of length 1\,808\,802 and 2\,923\,224.
The corresponding rates,
calculated including the acquisition time of the calibration data, source phase lock and basis switching, are~$\approx240$~bits/s and~$\approx~573$~bits/s.
If we consider only the time necessary for the data acquisition, we obtain net randomness rates of~$\approx 396$~bit/s and~$\approx 943$~bit/s.

These rates do not include the processing time of the Trevisan extractor.
This classical computation took 9 hours for \texttt{dataset1} and 22
days for \texttt{dataset2} on a machine processing 24 threads in parallel.
This task could be heavily parallelized. The bits extracted can be found in the ancillary files~\cite{ancillary}.

We used the NIST Statistical Test Suite~\cite{Rukhin:2010dn} to ensure the quality of generated strings is at least on par with acceptable pseudo-randomness. This suit of test can only verify the uniformity of the generated random string, it does not certify its privacy.
The string generated from \texttt{dataset1} passed all
the tests that are meaningful for this relatively short data sample, assuming an acceptable significance level~$\alpha=0.01$. The result of the individual tests are summarized in table~\ref{tab:NISTtest}.
\begin{table}[h]
    \centering
    \begin{tabular}{|l|c|c| }
        \hline
        \multicolumn{1}{|c|}{Test} & $P$--value & Proportion\\
        \hline
        Frequency & 0.590949 & 96/97\\
        Block Frequency & 0.275709 & 95/97\\
        Cumulative Sums Forward & 0.964295 & 96/97\\
        Cumulative Sums Backward & 0.637119 & 96/97\\
        Runs & 0.162606 & 97/97\\
        Longest Run of Ones & 0.590949 & 96/97\\
        Discrete Fourier Transform & 0.183769 & 96/97\\
        \hline
    \end{tabular}
    \caption{Result of the NIST Statistical Test Suite for the bits extracted from \texttt{dataset1}. We split the random bits into 97 sequences of 6300 bits each.}
    \label{tab:NISTtest}
\end{table}

\end{document}